\documentclass[11pt]{article}

\usepackage{amsmath}

\usepackage[cp866]{inputenc}

\begin{document}

\begin{center}

{\bf  V.M. Red'kov,  E.M. Ovsiuyk\\[3mm]
 TRANSITIVITY IN THE THEORY OF THE LORENTZ GROUP \\
 AND THE STOKES -- MUELLER FORMALISM IN  OPTICS}\\[3mm]
 Institute of Physics, National Academy of Sciences  of Belarus\\
 Mozyr State Pedagogical University\\
 redkov@dragon.bas-net.by;  e.ovsiyuk@mail.ru

\end{center}

%Report to be held:
%XV International School-Conference "Foundations  and  Advances in Nonlinear Science"
%September 20-23,   Minsk 2010

\begin{center}

{\bf Summary}

\end{center}

\begin{quotation}

Group-theoretical analysis of  arbitrary polarization devices is
performed, based on the theory of the Lorentz group. In effective
"non-relativistic" \hspace{2mm} Muel\-ler case, described by 3-dimensional orthogonal
matrices,  results of the one pola\-ri\-zation mea\-surement ${\bf S} \stackrel{O}{\rightarrow} {\bf S}'$
determine group theoretical parameters  within the accuracy of an arbitrary numerical variable.
There are derived formulas, defining Muller parameter of the non-relativistic Mueller device uniquely
 and in explicit form by by the results of two independent polarization measurements.

Analysis is extended to Lorentzian optical devices, described by
 4-dimen\-sional Mueller matrices.  In this case,  any single polarization measurement
 $(S_{0},{\bf S} )\stackrel{L}{\rightarrow} (S'_{0},{\bf S}')$ fixes parameters of the corresponding Mueller matrix
up to 3 arbitrary variables.
Formulas, defining Muller parameter of any relativistic Mueller device uniquely
can be  found  from results of four independent polari\-zation measurements.
Analytical  expressions for parameters of any Mueller device  can be  given the  most simple form
when using the results of 6 independent measurements,  the  corresponding formulas are written down in explicit form.

\end{quotation}

\subsection*{1. The transitivity problem in the theory of the Lorentz group
}

It is known that in describing (fully  or partly)   polarized  light
noticeable role may given to the group of $3+1$-pseudoorthog0nal transformations
consisting of a group $SO(3,1) $ isomorphic to the Lorentz group. Therefore,  techniques developed
 in the frames of the Lorentz group, in particular within relativistic kinematics, may play heuristic
  role in exploring  optical problems  (see  big list of references in the end;
a previous consideration of one of the authors is given in \cite{Redkov-2009}.

In the paper, when  working with the Lorentz group we  use technique  developed in
 \cite{Fedorov-1980} and \cite{Berezin-Kurochkin-Tolkachev-1989} and  partly updated  in \cite{Redkov-2009'}.
 This approach   had been started many years ago by Einstein and Mayer in
\cite{Einstein-Mayer(1)}.

Let us recall the known transitivity problem in relativistic kinematics:
in Stokes -- Mueller approach it reads
\begin{eqnarray}
L_{b}^{\;\;a}(k, \bar{k}^{*}) \;  S _{a} = + S' _{b} \;  .
\label{1}
\end{eqnarray}

\noindent From the very beginning,  one peculiarity shout be noted:
due to existence of the concept of little Lorentz group  initial and final Stokes 4-vectors  $S$ and $S'$,
one can write down  the transitivity condition in the form   $L \; (
L_{little} S ) = L'_{little} S'$,so that
\begin{eqnarray}
 [\; (L'_{little} )^{-1} L \;  L_{little} \; ]\; S   = S'
 \; .
 \label{2}
 \end{eqnarray}

\noindent This means that the transitive matrix $L$ cannot be defined uniquely  in terms of   $S$ and $S'$.

Let us use the factorized representation for Lorentzian matrices  (we adhere notation given in \cite{Redkov-2009, Redkov-2009'}),
eq.  (\ref{1}) gives
\begin{eqnarray}
A^{*} S = A^{-1} S' \; , \qquad \mbox{and} \qquad  A \;
S  =  ( A^{*})^{-1} \; S'  \; , \label{3}
\end{eqnarray}

\noindent or in more detailed form (conjugate  equation  is written  down too)
\begin{eqnarray}
\left |
\begin{array}{rrrr}
 k_{0}^{*}  & -k_{1}^{*}  &  -k_{2}^{*}  &  -k_{3}^{*}  \\
-k_{1}^{*}  &   k_{0}^{*}  & ik_{3}^{*}  &  -ik_{2}^{*}  \\
-k_{2}^{*} &  -ik_{3}^{*}  &   k_{0}^{*}  & ik_{1}^{*}  \\
-k_{3}^{*} & ik_{2}^{*}  &  -ik_{1}^{*}  &   k_{0}^{*}
\end{array}  \right |
\left | \begin{array}{c} S_{0} \\ S_{1} \\ S_{2} \\ S_{3}
\end{array} \right |=
 \left | \begin{array}{rrrr}
k_{0} & k_{1}  &  k_{2}  & k_{3}  \\
k_{1} & k_{0}  &  ik_{3}  & -ik_{2}  \\
k_{2} & -ik_{3}  &  k_{0}  & ik_{1}  \\
k_{3} & ik_{2}  &  -ik_{1}  & k_{0}
\end{array}  \right |
\left | \begin{array}{c} S'_{0} \\ S'_{1} \\ S'_{2} \\ S'_{3}
\end{array} \right |,
\nonumber
\end{eqnarray}
\begin{eqnarray}
\left |
\begin{array}{rrrr}
 k_{0}  & -k_{1}  &  -k_{2}  &  -k_{3}  \\
-k_{1} &   k_{0}  & -ik_{3}  &  ik_{2}  \\
-k_{2} &  ik_{3}  &   k_{0}  & -ik_{1}  \\
-k_{3} & -ik_{2}  &  ik_{1}  &   k_{0}
\end{array}  \right |
\left | \begin{array}{c} S_{0} \\ S_{1} \\ S_{2} \\ S_{3}
\end{array} \right |=
 \left | \begin{array}{rrrr}
k^{*}_{0} & k^{*}_{1}  &  k^{*}_{2}  & k^{*}_{3}  \\
k^{*}_{1} & k^{*}_{0}  &  -ik^{*}_{3}  & ik^{*}_{2}  \\
k^{*}_{2} & ik^{*}_{3}  &  k^{*}_{0}  & -ik^{*}_{1}  \\
k^{*}_{3} & -ik^{*}_{2}  &  ik^{*}_{1}  & k^{*}_{0}
\end{array}  \right |
\left | \begin{array}{c} S'_{0} \\ S'_{1} \\ S'_{2} \\ S'_{3}
\end{array} \right |.
\label{4}
\end{eqnarray}

\noindent
Below, the notation will be  used
\begin{eqnarray}
 k_{0} = n_{0} + i m_{0} \;, \qquad    k_{j} = -i  n_{j} +  m_{j} \;, \qquad k_{0}- {\bf k}^{2} = 1 \; .
 \nonumber
 \end{eqnarray}

\noindent  Summing and subtracting  eqs we get
\begin{eqnarray}
\left |
\begin{array}{rrrr}
 n_{0}  & -m_{1}  &  -m_{2}  &  -m_{3}  \\
-m_{1} &   n_{0}  & -n_{3}  &  n_{2}  \\
-m_{2} &  n_{3}  &   n_{0}  & -n_{1}  \\
-m_{3} & -n_{2}  &  n_{1}  &   n_{0}
\end{array}  \right |
\left | \begin{array}{c} S_{0} \\ S_{1} \\ S_{2} \\ S_{3}
\end{array} \right |=
 \left | \begin{array}{rrrr}
n_{0} & m_{1}  &  m_{2}  & m_{3}  \\
m_{1} & n_{0}  &  n_{3}  & -n_{2}  \\
m_{2} & -n_{3}  &  n_{0}  & n_{1}  \\
m_{3} & n_{2}  &  -n_{1}  & n_{0}
\end{array}  \right |
\left | \begin{array}{c} S'_{0} \\ S'_{1} \\ S'_{2} \\ S'_{3}
\end{array} \right | \; ,
\nonumber
\end{eqnarray}
\begin{eqnarray}
\left |
\begin{array}{rrrr}
-m_{0}  & -n_{1}  &  -n_{2}  &  -n_{3}  \\
-n_{1} &   -m_{0}  & m_{3}  &  -m_{2}  \\
-n_{2} &  -m_{3}  &   -m_{0}  & m_{1}  \\
-n_{3} & m_{2}  &  -m_{1}  &   -m_{0}
\end{array}  \right |
\left | \begin{array}{c} S_{0} \\ S_{1} \\ S_{2} \\ S_{3}
\end{array} \right |=
 \left | \begin{array}{rrrr}
m_{0} & -n_{1}  &  -n_{2}  & -n_{3}  \\
-n_{1} & m_{0}  &  m_{3}  & -m_{2}  \\
-n_{2} & -m_{3}  &  m_{0}  & m_{1}  \\
-n_{3} & m_{2}  &  -m_{1}  & m_{0}
\end{array}  \right |
\left | \begin{array}{c} S'_{0} \\ S'_{1} \\ S'_{2} \\ S'_{3}
\end{array} \right | \; . \;\; \;\;
\nonumber
\end{eqnarray}

\noindent So, we arrive at two homogeneous linear systems under 8 varianles
\begin{eqnarray}
n_{0} \; (S_{0} -S_{0}') - m_{1} \; (S_{1} + S'_{1} ) -  m_{2}\;
(S_{2} + S'_{2} ) -m_{3} \; (S_{3} + S'_{3} )  = 0 \; , \nonumber
\\
-m_{1} \; (S_{0} + S_{0}') + n_{0} \; (S_{1} -S_{1}') +n_{2} \;
(S_{3}  + S_{3}')  - n_{3} \; (S_{2} + S_{2}')  = 0 \; , \nonumber
\\
-m_{2} \; (S_{0} + S_{0}') + n_{0} \; (S_{2} -S_{2}')  + n_{3} \;
(S_{1}  + S_{1}')  - n_{1}\;  (S_{3} + S_{3}')   = 0\; , \nonumber
\\
-m_{3} \; (S_{0} + S_{0}') + n_{0}\;  (S_{3} -S_{3}')  + n_{1} \;
(S_{2}  + S_{2}')  - n_{2} \; (S_{1} + S_{1}')   = 0\; , \nonumber
\end{eqnarray}
\begin{eqnarray}
-m_{0} \; (S_{0} + S_{0}') - n_{1} \; (S_{1} - S'_{1} ) -  n_{2}
\; (S_{2} - S'_{2} ) -n_{3} \; (S_{3} - S'_{3} )  = 0 \; ,
\nonumber
\\
-n_{1} \; (S_{0} - S_{0}') - m_{0}\;  (S_{1} + S_{1}') -m_{2} \;
(S_{3}  - S_{3}')  + m_{3}\;  (S_{2} - S_{2}')  = 0 \; , \nonumber
\\
-n_{2} \; (S_{0} - S_{0}') - m_{0} \;  (S_{2} +S_{2}')  - m_{3}\;
(S_{1}  - S_{1}')  +m_{1} \; (S_{3} - S_{3}')   = 0\; , \nonumber
\\
-n_{3} \; (S_{0} - S_{0}') - m_{0} \; (S_{3} +S_{3}')  - m_{1} \;
(S_{2}
 - S_{2}')  + m_{2}\;  (S_{1} - S_{1}')   = 0\; .
\label{5}
\end{eqnarray}

\subsection*{2.  "Non-relativistic" \hspace{2mm} 3-dimensional Mueller matrices}

First, let us consider more simple (non-relativistic) case  when $S_{0}'=S_{0} = I = \mbox{inv}$.
Eqs.   (\ref{5} ) takes the form (because we search solutions in 3-dimensional rotations, we require  $m_{0}=0, \;  m_{j}=0$):
\begin{eqnarray}
 n_{0} \; (S_{1} -S_{1}') +n_{2} \; (S_{3}  + S_{3}')  - n_{3} \; (S_{2} + S_{2}')  = 0 \; ,
\nonumber
\\
 n_{0} \; (S_{2} -S_{2}')  + n_{3} \; (S_{1}  + S_{1}')  - n_{1}\;  (S_{3} + S_{3}')   = 0\; ,
\nonumber
\\
 n_{0}\;  (S_{3} -S_{3}')  + n_{1} \; (S_{2}  + S_{2}')  - n_{2} \; (S_{1} + S_{1}')   = 0\; ,
\nonumber
\\
- n_{1} \; (S_{1} - S'_{1} ) -  n_{2} \; (S_{2} - S'_{2} ) -n_{3}
\; (S_{3} - S'_{3} )  = 0 \; . \label{6}
\end{eqnarray}

\noindent The fourth equation in not independent of three remaining -- it follows from them.
Therefore we have  the system of 3 independent ones
\begin{eqnarray}
 n_{2} \; (S_{3}  + S_{3}')  - n_{3} \; (S_{2} + S_{2}')  =  - n_{0} \; (S_{1} -S_{1}')  \; ,
\nonumber
\\
 n_{3} \; (S_{1}  + S_{1}')  - n_{1}\;  (S_{3} + S_{3}')   = - n_{0} \; (S_{2} -S_{2}')\; ,
\nonumber
\\
 n_{1} \; (S_{2}  + S_{2}')  - n_{2} \; (S_{1} + S_{1}')   = -  n_{0}\;  (S_{3} -S_{3}')\; .
\label{7}
\end{eqnarray}

\noindent They may be written  in 3-vector form
\begin{eqnarray}
{\bf n} \times ( {\bf S} + {\bf S}') = -n_{0} \; ( {\bf S} - {\bf
S}') \; .
\label{8}
\end{eqnarray}

\noindent
General solutions for ${\bf n}$ can be searched with the aid of substitution
 \begin{eqnarray}
  {\bf n} = \alpha \; {\bf S} + \rho \; {\bf S} ' +
\beta \; {\bf S} \times {\bf S}'\;,
\nonumber
\end{eqnarray}
 then eq. (\ref{8}) leads to  (below note  $S^{2} = {\bf S} {\bf S}$)
\begin{eqnarray}
(\alpha - \rho  )  \; {\bf S} \times {\bf S}' + \beta \; [ \;
{\bf S}'\; S^{2}  + {\bf S}'\; ({\bf S} {\bf S}') - {\bf S}\;
S^{2} -  {\bf S}\; ({\bf S} {\bf S}') \; ] =   -n_{0}  {\bf S}    + n_{0} {\bf S}'\; ,
\nonumber
\end{eqnarray}

\noindent from whence it follow  $\rho  = \alpha \; ,\; \alpha$ is arbitrary, and
\begin{eqnarray}
 n_{0} =  \beta \; ( S^{2} +
{\bf S} \; {\bf S}' ) \; ,\qquad
{\bf n} = \alpha \; ( {\bf S} +  {\bf S} '  ) + \beta \;    {\bf S} \times {\bf S}'   \;
  .
\label{9}
\end{eqnarray}

One  must to take into account
additional restriction for parameters of rotation matrices
\begin{eqnarray}
 n_{0}^{2}  + {\bf n}^{2} = 1\; , \qquad
\label{11}
\end{eqnarray}

\noindent which results in
\begin{eqnarray}
 \beta ^{2}\; ( S^{2} +
{\bf S} \; {\bf S}' )^{2}  + [ \alpha \; ( {\bf S} +  {\bf S} '  ) + \beta   \;  {\bf
S} \times {\bf S}'  ]^{2} = 1\; ,
\nonumber
\end{eqnarray}

\noindent
or
\begin{eqnarray}
 \beta ^{2}\; [\;  S^{4} + 2 S^{2} \; ({\bf S} \; {\bf S}') + ({\bf S} \; {\bf S}')^{2} ]
+
\beta ^{2}  \; [
S^{4}  - ({\bf S}  {\bf S}' ) ^{2}  ]   +
  2 \alpha^{2}  \; ( S^{2}  +  {\bf S}\; {\bf S}')   = 1 \; ;
\nonumber
\end{eqnarray}

\noindent
and ultimately eq. (\ref{11}) gives
\begin{eqnarray}
 \beta ^{2} \;  S^{2}\;  +
   \alpha^{2}     = {1 \over 2( S^{2}  +  {\bf S}\; {\bf S}')}\; .
\label{12}
\end{eqnarray}

\noindent
General solution of eq. (\ref{12}) can be presented in terms of $\sin$- and $\cos$-functions  of an angular variable
\begin{eqnarray}
\alpha = {\sin \Gamma \over   \sqrt{2(S^{2}  +  {\bf S}\; {\bf S}')}}\; , \qquad
\beta = {\cos \Gamma \over   S\; \sqrt{2(S^{2}  +  {\bf S}\; {\bf S}')}} \; , \qquad
\Gamma \in [0, 2\pi ]\; .
\label{13}
\end{eqnarray}

\noindent
Thus,  relations  (\ref{9}) read (here  $\Gamma \in [0 , 2 \pi ]$ stands for arbitrary  parameter)
\begin{eqnarray}
 n_{0}^{2}  + {\bf n}^{2} = 1\; ,
\qquad
 n_{0} =  {\cos \Gamma \over   S\; \sqrt{2(S^{2}  +  {\bf S}\; {\bf S}')}}  \; ( S^{2} +
{\bf S} \; {\bf S}' ) \; ,
\nonumber
\\
{\bf n} = {\sin \Gamma \over   \sqrt{2(S^{2}  +  {\bf S}\; {\bf S}')}} \; ( {\bf S} +  {\bf S} '  ) +  {\cos \Gamma \over   S\; \sqrt{2(S^{2}  +  {\bf S}\; {\bf S}')}}  \;    {\bf
S} \times {\bf S}'   \;
.
\label{14}
\end{eqnarray}

Note that when  ${\bf S}' = {\bf S}$, relations (\ref{14}) describe the case of little rotation group
\begin{eqnarray}
 n_{0}^{2}  + {\bf n}^{2} = 1\; ,
\qquad
 n_{0} =  \cos \Gamma  \; , \qquad
{\bf n} = \sin \Gamma \; { {\bf \bf S} \over S }\; .
\label{15}
\end{eqnarray}

When $\Gamma =0 $ , solution  (\ref{14})  becomes of  the  most simple form
\begin{eqnarray}
 n_{0} =  { S^{2} + {\bf S} \; {\bf S}'   \over
  S\; \sqrt{2 \; (S^{2}  +  {\bf S}\; {\bf S}')}}  \;  \; ,\qquad
{\bf n} =    { {\bf S} \times {\bf S}'  \over   S\; \sqrt{2 \; (S^{2}  +  {\bf S}\; {\bf S}')}}  \; .
\label{16}
\end{eqnarray}

Note, that we may transform all the relations to a Gibbs 3-vector parameter
in the rotation group  (the full treatment of the  theory in this parametrization see in \cite{Fedorov-1980})
\begin{eqnarray}
{\bf c} = { {\bf n} \over n_{0} } \; ,
\label{c}
\end{eqnarray}

\noindent  then eqs.  (\ref{14}) give
\begin{eqnarray}
{\bf c} = \; \mbox{tg}\;  \Gamma \; { S\over   S^{2} +{\bf S} \; {\bf S}'    } \; ( {\bf S} +  {\bf S} '  ) +
 { {\bf S} \times {\bf S}'    \over     S^{2} +{\bf S} \; {\bf S}'  }  \;      .
\label{17}
\end{eqnarray}

Note that in the non-relativistic case, for Stokes vectors one can use the following parametrization
($I$ is intensity of the light beam, $p$ is a polarization degree)
\begin{eqnarray}
S_{0} = I, \qquad {\bf S} =  I p  \; {\bf N} , \qquad I - \mbox{inv} \; , \qquad  {\bf N}^{2} = 1 \; ;
\label{18}
\end{eqnarray}

\noindent
at this  (\ref{14}) and (\ref{16}) change to
\begin{eqnarray}
 n_{0}^{2}  + {\bf n}^{2} = 1\; ,
\qquad
 n_{0} =  \cos \Gamma  {  1 + {\bf N} \; {\bf N}'  \over   \sqrt{2( 1  +  {\bf N}\; {\bf N}')}}  \;
  \; ,
\nonumber
\\
{\bf n} = \sin \Gamma \; {    {\bf N} +  {\bf N} '   \over   \sqrt{2(1  +  {\bf N}\; {\bf N}')}}
\; +  \; \cos \Gamma\; {  {\bf N} \times {\bf N}' \over    \sqrt{2(1  +  {\bf N}  {\bf N}')}}  \;       \;
,
\label{19}
\end{eqnarray}

\noindent and
\begin{eqnarray}
{\bf c} = \; \mbox{tg}\;  \Gamma \; {   {\bf N} +  {\bf N} '   \over   1  +{\bf N} \; {\bf N}'    } \;  +
 { {\bf N} \times {\bf N}'    \over     1  +{\bf N} \; {\bf N}'  }  \;      \;
.
\label{20}
\end{eqnarray}

\subsection*{3.  On defining  Mueller 3-matrices from the results of
polarization measurements}

Because a single polarization measurement relating  $ {\bf S} \stackrel{L}{\longrightarrow } {\bf S}_{1}'$
cannot fix  Mueller 3-matrix uniquely, to obtain result values for parameters of the Mueller 3-matrix,
one need to perform two independent measurements
${\bf S}_{1}  \stackrel{L}{\longrightarrow } {\bf S}_{1}', \;\;  {\bf S}_{2}  \stackrel{L}{\longrightarrow } {\bf S}_{2}'$.
Mathematically,  the problem of finding a definite Mueller 3-matrix can be  formulated as
a system to solve, describing two polarization measurement with one the same Mueller matrix.

First, let us consider this task  with the aid of Gibbs 3-paramere
\begin{eqnarray}
{\bf c} = \; \mbox{tg}\;  \Gamma \; {   {\bf N}_{1} +  {\bf N}_{1} '   \over   1  +{\bf N}_{1} {\bf N}'_{1}    } \;  +
 { {\bf N}_{1} \times {\bf N}' _{1}   \over     1  +{\bf N}_{1} {\bf N}'_{1}  } \;, \;\;
 {\bf c} =
  \; \mbox{tg}\;  \Gamma \; {   {\bf N}_{2} +  {\bf N}_{2} '   \over   1  +{\bf N}_{2} {\bf N}'_{2}    } \;  +
 { {\bf N}_{2} \times {\bf N}' _{2}   \over     1  +{\bf N}_{2} {\bf N}' _{2} }  \;      ;
\label{21}
\end{eqnarray}

\noindent
so we have a vector equation
\begin{eqnarray}
\mbox{tg}\;  \Gamma \; \left [ {   {\bf N}_{1} +  {\bf N}_{1} '   \over   1  +{\bf N}_{1} {\bf N}'_{1}    }
-  {   {\bf N}_{2} +  {\bf N}_{2} '   \over   1  +{\bf N}_{2} {\bf N}'_{2}    } \right ] +
 { {\bf N}_{1} \times {\bf N}' _{1}   \over     1  +{\bf N}_{1} {\bf N}'_{1}  } -
{ {\bf N}_{2} \times {\bf N}' _{2}   \over     1  +{\bf N}_{2} {\bf N}' _{2} } = 0\; .
\label{22}
\end{eqnarray}

\noindent
Multiplying it by  ${\bf N}_{1}, {\bf N}_{1}', {\bf N}_{2},{\bf N}_{2}'$, we obtain four scalar equations
\begin{eqnarray}
\mbox{tg}\;  \Gamma   \left [ \;  1
-  {  {\bf N}_{1}( {\bf N}_{2} +  {\bf N}_{2} ' )  \over   1  +{\bf N}_{2} {\bf N}'_{2}    } \; \right ] -
{ {\bf N}_{1}({\bf N}_{2} \times {\bf N}' _{2} )  \over     1  +{\bf N}_{2} {\bf N}' _{2} } =0 \; ,
\nonumber
\\
\mbox{tg}\;  \Gamma  \left [ \;  1
-  {  {\bf N}_{1}' ( {\bf N}_{2} +  {\bf N}_{2} '  ) \over   1  +{\bf N}_{2} {\bf N}'_{2}    } \;  \right ]
 -
{ {\bf N}_{1}' ({\bf N}_{2} \times {\bf N}' _{2} )  \over     1  +{\bf N}_{2} {\bf N}' _{2} } = 0\; ,
\nonumber
\\
\mbox{tg}  \Gamma  \left [ \; { {\bf N}_{2}(  {\bf N}_{1} +  {\bf N}_{1} ' )  \over   1  +{\bf N}_{1} {\bf N}'_{1}    }
-  1  \; \right ] +
 { {\bf N}_{2}({\bf N}_{1} \times {\bf N}' _{1} )  \over     1  +{\bf N}_{1} {\bf N}'_{1}  } =0 \; ,
\nonumber
\\
\mbox{tg}  \Gamma  \left [ \;  { {\bf N}_{2}'(  {\bf N}_{1} +  {\bf N}_{1} '  ) \over   1  +{\bf N}_{1} {\bf N}'_{1}    }
-  1 \; \right ] \; + \;
 { {\bf N}_{2}' {\bf N}_{1} \times {\bf N}' _{1}   \over     1  +{\bf N}_{1} {\bf N}'_{1}  } = 0 \; .\;
\label{23}
\end{eqnarray}

\noindent
From whence it follow
\begin{eqnarray}
\mbox{tg}\;  \Gamma \;   =
  { {\bf N}_{1} \; ({\bf N}_{2} \times {\bf N}' _{2} )  \over
 ( {\bf N}_{2} -  {\bf N}_{1}  ) ( {\bf N}_{2} +  {\bf N}_{2} ' ) }    \; , \qquad
\mbox{tg}\;  \Gamma \;
 = -
{ {\bf N}_{1}' \; ( {\bf N}' _{2}  \times  {\bf N}_{2}  )  \over
( {\bf N}_{2}' -  {\bf N}_{1}'  ) ( {\bf N}_{2}' +  {\bf N}_{2}  )   } \; ,
\nonumber
\\
\mbox{tg}\;  \Gamma \;  =
  {  {\bf N}_{2} \; ({\bf N}_{1} \times {\bf N}' _{1} )  \over
 (  {\bf N}_{1} -  {\bf N}_{2}  )  (  {\bf N}_{1} +  {\bf N}_{1} ' )        }    \; ,
\qquad
\mbox{tg}\;  \Gamma
 = -
 { {\bf N}_{2}'\; ( {\bf N}' _{1}  \times  {\bf N}_{1}  )  \over
  (  {\bf N}_{1}' -  {\bf N}_{2}'  )  (  {\bf N}_{1}' +  {\bf N}_{1}  )        }  \; .
 \label{24}
\end{eqnarray}

Thus, we have a  simple expression for  $ \mbox{tg} \; \Gamma \;$,
together with four additional constraints, which determine the whole  aggregate of all possible couples of Stokes 3-vectors
related by one the same Mueller matrices.

Now let us  detail considering of the task in
the frames of unitary group $SU(2)$ -- evidently,  two solutions   cannot contradict each other.
Here we have
\begin{eqnarray}
 n_{0} =  \beta_{1} \;  {\bf S}_{1} (  {\bf S}_{1}  +  {\bf S}'_{1} ) \; ,\qquad
{\bf n} = \alpha _{1}\; ( {\bf S}_{1} +  {\bf S} '_{1}  ) + \beta_{1} \;    {\bf S} _{1}\times {\bf S}'_{1}   \; ,
\nonumber
\\
 n_{0} =  \beta_{2} \; {\bf S}_{2} (  {\bf S}_{2}  +  {\bf S}'_{2} ) \; ,\qquad
{\bf n} = \alpha _{2}\; ( {\bf S}_{2} +  {\bf S} '_{2}  ) + \beta_{2} \;    {\bf S} _{2}\times {\bf S}'_{2}   \;  .
\label{25}
\end{eqnarray}

\noindent
what is equivalent to
\begin{eqnarray}
n_{0} = \cos \Gamma \; {1  +  {\bf N}_{1} {\bf N}_{1}' \over  \sqrt{2 (1 + {\bf N}_{1} {\bf N}_{1}')}}\; , \qquad
{\bf n} = \sin \Gamma \;
{  {\bf N}_{1}  + {\bf N}_{1}' \over  \sqrt{2 (1 + {\bf N}_{1} {\bf N}_{1}')}} +  \cos \Gamma\;
{ {\bf N}_{1} \times {\bf N}'_{1} \over \sqrt{2 (1 + {\bf N}_{1} {\bf N}_{1}')}}
\nonumber
\\
n_{0} = \cos \Gamma \; {1  +  {\bf N}_{2} {\bf N}_{2}' \over  \sqrt{2 (1 + {\bf N}_{2} {\bf N}_{2}')}}\; , \qquad
{\bf n} = \sin \Gamma \;
{  {\bf N}_{2}  + {\bf N}_{2}' \over  \sqrt{2 (1 + {\bf N}_{2} {\bf N}_{2}')}} +  \cos \Gamma\;
{ {\bf N}_{2} \times {\bf N}'_{2} \over \sqrt{2 (1 + {\bf N}_{2} {\bf N}_{2}')}}
\nonumber
\\
\label{26}
\end{eqnarray}

From two different  expressions for $n_{0}$, it follows
\begin{eqnarray}
{\bf N}_{1} {\bf N}_{1}' = {\bf N}_{2} {\bf N}_{2}' \; .
\label{27}
\end{eqnarray}

\noindent
Taking this into account, from two different expressions for  ${\bf n}$ we derive
\begin{eqnarray}
\sin \Gamma \;[\;
 (  {\bf N}_{1}  + {\bf N}_{1}' ) -  (  {\bf N}_{2}  + {\bf N}_{2}' ) \; ] +  \cos \Gamma\;
[\; ( {\bf N}_{1} \times {\bf N}'_{1} ) - ( {\bf N}_{2} \times {\bf N}'_{2} ) \ ] =0
\label{28}
\end{eqnarray}

It should be noted that due to (\ref{27}), relation  (\ref{22}) becomes much more simpler
\begin{eqnarray}
\mbox{tg}\;  \Gamma \; [ \;    ( {\bf N}_{1} +  {\bf N}_{1} ' ) -    ( {\bf N}_{2} +  {\bf N}_{2} ') \; ] +
  {\bf N}_{1} \times {\bf N}' _{1}   -
 {\bf N}_{2} \times {\bf N}' _{2}   = 0\; .
\label{29}
\end{eqnarray}

\noindent
In fact, (\ref{28}) and (\ref{29}) coincide, difference  consist in the following:
 (\ref{28}) cannot distinguish between two solutions:
$
(+ \cos \Gamma, \; + \sin \Gamma )$ and $ (- \cos \Gamma, \; - \sin \Gamma )$.

\subsection*{4. Relativistic Mueller matrices relating two Stokes 4-vectors }

Let us turn back to general (relativistic) case of Mueller matrices (\ref{5}):
\begin{eqnarray}
  m_{1} \; (S_{1} + S'_{1} ) +  m_{2}\; (S_{2} + S'_{2} ) + m_{3} \; (S_{3} + S'_{3} )  =  n_{0} \; (S_{0} -S_{0}') \; ,
\nonumber
\\
m_{1} \; (S_{0} + S_{0}')   - n_{2} \; (S_{3}  + S_{3}')  + n_{3}
\; (S_{2} + S_{2}')  =   n_{0} \; (S_{1} -S_{1}') \; , \nonumber
\\
m_{2} \; (S_{0} + S_{0}')   - n_{3} \; (S_{1}  + S_{1}')  +
n_{1}\;  (S_{3} + S_{3}')   =  n_{0} \; (S_{2} -S_{2}')\; ,
\nonumber
\\
m_{3} \; (S_{0} + S_{0}')   - n_{1} \; (S_{2}  + S_{2}')  + n_{2}
\; (S_{1} + S_{1}')   =   n_{0}\;  (S_{3} -S_{3}')\; , \nonumber
\end{eqnarray}
\begin{eqnarray}
 - n_{1} \; (S_{1} - S'_{1} ) -  n_{2} \; (S_{2} - S'_{2} ) -n_{3} \; (S_{3} - S'_{3} )  = m_{0} \; (S_{0} + S_{0}') \; ,
\nonumber
\\
-n_{1} \; (S_{0} - S_{0}') -m_{2} \; (S_{3}  - S_{3}')  + m_{3}\;
(S_{2} - S_{2}')  =  m_{0}\;  (S_{1} + S_{1}')  \; , \nonumber
\\
-n_{2} \; (S_{0} - S_{0}')   - m_{3}\;  (S_{1}  - S_{1}')  +m_{1}
\; (S_{3} - S_{3}')   =  m_{0} \;  (S_{2} +S_{2}')\; , \nonumber
\\
-n_{3} \; (S_{0} - S_{0}')   - m_{1} \;  (S_{2}
 - S_{2}')  + m_{2}\;  (S_{1} - S_{1}')   =  m_{0} \; (S_{3} +S_{3}') \; .
\label{31}
\end{eqnarray}

\noindent Because we  search solutions among proper orthochronous Lorentzian  transformations,
 unknown parameters  must obey additional relations
\begin{eqnarray}
n_{0}^{2}  + {\bf n} ^{2}  - m_{0}^{2}  - {\bf m} ^{2} = 1 \; ,
\qquad n_{0} m_{0} + {\bf n} {\bf m} = 0 \; ; \label{32}
\end{eqnarray}

\noindent by this reason, the trivial solution  $n_{a}=0 ,
m_{a}=0$  for    (\ref{31}) is of no interest. Eqs.  (\ref{31}) can be rewritten in 3-vector  form
\begin{eqnarray}
 {\bf m}\;  ( {\bf S} + {\bf S}') = n_{0}
\; (S_{0} - S'_{0}) \; ,
\nonumber \\
 {\bf n}\;  ( {\bf S} - {\bf
S}') = -m_{0} \; (S_{0} + S'_{0}) \; , \nonumber
\\
{\bf m} \; (S_{0} + S'_{0}) + ( {\bf S} + {\bf S}')
\times {\bf n} = n_{0} \; ( {\bf S} - {\bf S}') \; , \nonumber
\\
{\bf n} \; (S_{0} - S'_{0}) -  ( {\bf S} - {\bf S}')
\times {\bf m } = - m_{0} \; ( {\bf S} + {\bf S}') \; .
\label{33}
\end{eqnarray}

\noindent Note that the (non-relativity) requirement  $S_{0} - S'_{0} = 0$ immediately
leads us to additional  relations ${\bf m}= 0$ and  $m_{0} =0$, and we get eqs.  (\ref{7})--(\ref{8}).

Let us introduce notation
\begin{eqnarray}
 S_{0} + S'_{0}= A \;, \qquad  S_{0} - S'_{0}= B \;, \qquad
 {\bf S} + {\bf S}'  = {\bf A} \; ,
\nonumber
\\
  {\bf S} - {\bf S}' = {\bf B}\; ,\qquad
N_{+} = \nu, \qquad  M_{-} = \mu \; ;
\label{32}
\end{eqnarray}

The  complete  system od equations to solve  is
\begin{eqnarray}
n_{0}^{2}  + {\bf n} ^{2}  - m_{0}^{2}  - {\bf m} ^{2} = 1 \; ,
\qquad n_{0} m_{0} + {\bf n} {\bf m} = 0 \; ;
\label{33}
\\
 {\bf m}\;   {\bf A} = n_{0}
\; B  \; , \qquad
 {\bf n}\;   {\bf B}  = -m_{0} \; A  \; ;
\label{34}
\\
{\bf m} \; A  +  {\bf A} \times {\bf n} = n_{0} \;  {\bf B} \; , \qquad
{\bf n} \; B -   {\bf B} \times {\bf m } = - m_{0} \;  {\bf A}  \; .
\label{35}
\end{eqnarray}

In is convenient to use  linear expansions for both 3-vectors
\begin{eqnarray}
{\bf n} = N_{+}   {\bf A}  + N_{-}  {\bf B} + N
 {\bf A}  \times  {\bf B} \; , \qquad
{\bf m} =  M_{+}  {\bf A}  + M_{-}  {\bf B}  + M
 {\bf A}  \times  {\bf B}   \; .
\label{36}
\end{eqnarray}

From the first equation in (\ref{35}) it follows
\begin{eqnarray}
A ( M_{+}  {\bf A}  + M_{-} {\bf B}  + M
 {\bf A} \times  {\bf B}  ) +
   {\bf A}  \times (  N_{-}  {\bf B}  + N
 {\bf A}  \times  {\bf B} )
=  n_{0} \;  {\bf B}  \; ,
\nonumber
\end{eqnarray}

\noindent
which gives three equations
\begin{eqnarray}
A  M_{+} +  {\bf A} {\bf B}\;  N = 0 \; , \qquad
A  M_{-} -   {\bf A} ^{2}    N =  n_{0} \; , \qquad
A  M + N_{-} = 0 \; .
\label{37}
\end{eqnarray}

\noindent
In the same manner, from the second equation in  (\ref{35}) we get
\begin{eqnarray}
 \; B \; ( N_{+}   {\bf A}  + N_{-}  {\bf B} + N
 {\bf A}  \times  {\bf B} )
 -
   {\bf B}
\times  (
M_{+}  {\bf A}   + M
 {\bf A} \times  {\bf B} )
 = - m_{0} \;  {\bf A}\; ,
\nonumber
\end{eqnarray}

\noindent
and further
\begin{eqnarray}
B  N_{-}  +   {\bf A}  {\bf B}     M =  0  \; , \qquad
B N_{+} -   {\bf B} ^{2}\;  M = -m_{0}  \; ,
\qquad
B  N + M_{+} = 0 \; .
\label{38}
\end{eqnarray}

Thus, two  vector equations (\ref{35}) provide us with the system for
six parameters
\begin{eqnarray}
A  M_{+} +  {\bf A} {\bf B}\;  N = 0 \; , \qquad
A  M_{-} -   {\bf A} ^{2}    N =  n_{0} \; , \qquad
A  M + N_{-} = 0 \; ;
\nonumber
\\
B  N_{-}  +   {\bf A} {\bf B}     M =  0  \; , \qquad
B  N_{+} -   {\bf B}  ^{2}\;  M = -m_{0}  \; , \qquad
B  N + M_{+} = 0 \; .
\label{39}
\end{eqnarray}

\noindent After excluding the variables  $N_{-},M_{+}$:
\begin{eqnarray}
 N_{-} = - A  M \;  , \qquad
 M_{+}= - B  N  \; ,
\label{40}
\end{eqnarray}

\noindent eqs. (\ref{39}) read
\begin{eqnarray}
- AB   N +  {\bf A}{\bf B} \;  N = 0 \; , \qquad
A  M_{-} -   {\bf A}  ^{2}    N =  n_{0} \; ,
\nonumber
\\
-AB  M  +   {\bf A} {\bf B}     M =  0  \; ,
\qquad
B  N_{+} -   {\bf B} ^{2}\;  M = -m_{0}  \; .
\label{41}
\end{eqnarray}

\noindent
Note that equations  1 and 3 are identities.
In fact, eqs.  (\ref{41}) are equivalent to two equations only
\begin{eqnarray}
A  M_{-} -   {\bf A}  ^{2}    N =  n_{0} \; , \qquad
B  N_{+} -   {\bf B} ^{2}\;  M = -m_{0}  \; ,
\label{42}
\end{eqnarray}

Substituting expressions
\begin{eqnarray}
{\bf n} = N_{+}   {\bf A} -   M \; A    {\bf B} + N  {\bf A}  \times  {\bf B}   \; , \qquad
{\bf m} =  M_{-}  {\bf B}    - N \; B    {\bf A}  + M  {\bf A} \times  {\bf B}   \; ;
\label{43}
\end{eqnarray}

\noindent into  (\ref{34}),  we  arrive at
\begin{eqnarray}
M_{-}  {\bf B}{\bf A}    - N \; B    {\bf A} ^{2}  = n_{0} \; B  \;
\;\; \Longrightarrow \;\; M_{-}   A    - N   {\bf A} ^{2}  = n_{0}  \; ,
\nonumber
\\
 N_{+}   {\bf A} {\bf B} -   M \; A    {\bf B}^{2}  = -m_{0} \; A  \; \;\; \Longrightarrow \;\;
 N_{+}    B -   M \;   {\bf B}^{2}  = -m_{0} \; ;
  \nonumber
\end{eqnarray}

\noindent  which coincide with (\ref{42}). This means that eqs. (\ref{34}) can be removed.
The above substitutions for two vectors  (\ref{43}) are to be allowed in the conditions
\begin{eqnarray}
n_{0}^{2}  - m_{0}^{2}  = 1 + {\bf m} ^{2} -  {\bf n} ^{2}    \; ,
\qquad n_{0} m_{0} =-  {\bf n} {\bf m} = 0 \; .
\nonumber
\end{eqnarray}

Let us simplify notation
$$
M_{-} = x \; , \qquad N = y, \qquad N_{+} = z\;, \qquad M= w
$$

\noindent
In these variables, the main equations to solve read
\begin{eqnarray}
  n_{0}  = A \; x  -   {\bf A} ^{2}    y \; , \qquad
  {\bf n} = z   {\bf A}   -   w  A  {\bf B}  + y {\bf A} \times {\bf B} |; ,
\nonumber
\\
m_{0}  = - B \; z  +   {\bf B}^{2}\;  w  \; ,
\qquad
{\bf m} =  x {\bf B}    - y \; B \; {\bf A}   + w
{\bf A} \times {\bf B}  \; ;
\nonumber
\\
n_{0} m_{0} = -  {\bf n} {\bf m}  \; ,
\qquad
n_{0}^{2}  - m_{0}^{2}  = 1 + {\bf m} ^{2} -  {\bf n} ^{2}    \; .
\label{45}
\end{eqnarray}

First, let us detail $n_{0} m_{0} = -  {\bf n} {\bf m}$. Taking into account
\begin{eqnarray}
n_{0} m_{0}=
-x z \;  AB  + w x \; A {\bf B}^{2}\;  + y z \;  B {\bf A} ^{2}-
wy \; {\bf A} ^{2}    {\bf B}^{2}\; ,
\nonumber
\\
- {\bf n} {\bf m}= -(z   {\bf A}   -   w  A  {\bf B}  + y {\bf A} \times {\bf B})\,(x {\bf B}
  - y \; B \; {\bf A}   + w
{\bf A} \times {\bf B})=
\nonumber
\\
= - x z \;   {\bf A} {\bf B} +  y   z \;  B {\bf A} ^{2}  + w x \;  A {\bf B}^{2} -
 yw \;   AB  {\bf A} {\bf B} -  yw
{\bf A} ^{2}    {\bf B}^{2}  +   yw  (  {\bf A}{\bf B})^{2} \; .
\nonumber
\end{eqnarray}

\noindent
we arrive at
\begin{eqnarray}
0 = x z \; ( AB -   {\bf A} {\bf B} )
 -
 yw  \;   AB  {\bf A} {\bf B} +   yw  (  {\bf A}{\bf B})^{2}  \; .
\label{46}
\end{eqnarray}

\noindent
Because
\begin{eqnarray}
AB - {\bf A} {\bf B} = (S_{0}^{2} - {\bf S}^{2}) -  (S_{0}^{'2} - {\bf S}^{'2})
= 0 \; ,
\label{47}
\end{eqnarray}

\noindent eq. (\ref{46})  takes the form of an identity  $0 = 0$,
 subsequently, this equation can be excluded  from
(\ref{45}). Remaining and independent relations are
\begin{eqnarray}
n_{0}^{2}  - m_{0}^{2}  = 1 + {\bf m} ^{2} -  {\bf n} ^{2}    \; ,
\nonumber
\\
  n_{0}  = A \; x  -   {\bf A} ^{2}    y \; , \qquad
  {\bf n} = z   {\bf A}   -   w  A  {\bf B}  + y {\bf A} \times {\bf B} \; ,
\nonumber
\\
m_{0}  = - B \; z  +   {\bf B}^{2}\;  w  \; ,
\qquad
{\bf m} =  x {\bf B}    - y \; B \; {\bf A}   + w
{\bf A} \times {\bf B}  \; .
\label{48}
\end{eqnarray}

Each of vector equation in (\ref{48})  can be changed into three scalar ones; those
are obtained through  multiplying  them by  ${\bf A}, {\bf B} , {\bf A} \times {\bf B}$:
\begin{eqnarray}
 {\bf A}  {\bf n} = z  \;  {\bf A}^{2}   -   w  \; A ^{2} B  \; ,
\nonumber
\\
{\bf B} {\bf n} = z  \;  AB   -   w  \; A  {\bf B} ^{2}  \; ,
\nonumber
\\
( {\bf A} \times {\bf B}) {\bf n} = + y \; {\bf A}^{2}  {\bf B}^{2} -y\;  A^{2} B^{2}  \; ,
\nonumber
\\[3mm]
{\bf A} {\bf m} =  x \; AB     - y \; B \; {\bf A} ^{2}
\nonumber
\\
{\bf B}  {\bf m} =  x \; {\bf B}^{2}     - y \; B ^{2} A\; ,
\nonumber
\\
( {\bf A} \times {\bf B}) {\bf m} =
+ w \; {\bf A}^{2}  {\bf B}^{2} - w\;  A^{2} B^{2}  \;
 .
\label{49}
\end{eqnarray}

These equations are  easy to solve
\begin{eqnarray}
y =  {( {\bf A} \times {\bf B}) {\bf n} \over  {\bf A}^{2}  {\bf B}^{2} -  A^{2} B^{2}  }\;,\;
z = - { ({\bf B} {\bf n} ) AB - ({\bf A} {\bf n}) {\bf B}^{2}  \over  {\bf A}^{2}  {\bf B}^{2} -  A^{2} B^{2}   }\;,\;
w = - {1 \over A} { ({\bf B} {\bf n} ) {\bf A}^{2} - ({\bf A} {\bf n}) AB  \over
  {\bf A}^{2}  {\bf B}^{2} -  A^{2} B^{2}   }\; ;
\nonumber
\\
w =  {( {\bf A} \times {\bf B}) {\bf m} \over  {\bf A}^{2}  {\bf B}^{2} -  A^{2} B^{2}  }\;,\;
x =  { - ({\bf A} {\bf m} ) AB +  ({\bf B} {\bf m}) {\bf A}^{2}  \over  {\bf A}^{2}  {\bf B}^{2} -  A^{2} B^{2}   }\;,\;
y =  {1 \over B} { ({\bf B} {\bf m} )
 AB  - ({\bf A} {\bf m}) {\bf B}^{2} \over  {\bf A}^{2}  {\bf B}^{2} -  A^{2} B^{2}   }\;.
\nonumber
\\
\label{50}
\end{eqnarray}

Taking  (\ref{48}),  we  may turn back to a starting complex parameter   $k_{a}$:
\begin{eqnarray}
k_{0} = n_{0} + i m_{0}= (  x A   -i z  B )  -   (y {\bf A} ^{2}  -  i w {\bf B}^{2} ) \; ,
\nonumber
\\
{\bf k} = {\bf m} - i {\bf n} =
  - ( y \; B   + i \;  z  ) \; {\bf A}  +
   ( x    +i    w  A ) \;  {\bf B}    +
   ( w      -i y ) {\bf A} \times {\bf B} \; .
\label{51}
\end{eqnarray}

\noindent Note that one can derive a more simple 3-vector, parameter for Lorentz group [...],
\begin{eqnarray}
{\bf q} = {{\bf k} \over  k_{0}} =
 {  -  ( y \; B   + i \;  z  ) \; {\bf A}  +
   ( x    +i    w  A ) \;  {\bf B}    +
   ( w      -i y ) {\bf A} \times {\bf B}   \over
(x A  -  i z  B )    - (  y {\bf A} ^{2}       - i  w {\bf B}^{2} )  }
\label{52}
\end{eqnarray}

\noindent
It may be formally simplified
\begin{eqnarray}
{\bf q} = \alpha \; {\bf A} + \beta \; {\bf B} + \gamma \; {\bf A} \times {\bf B} \; ,
\nonumber
\\
\alpha = { - ( y \; B   + i \;  z)   \over
(x A  -  i z  B )    - (  y {\bf A} ^{2}       - i  w {\bf B}^{2} )} \; ,
\nonumber
\\
\beta = {   x    +i    w  A  \over
(x A  -  i z  B )    - (  y {\bf A} ^{2}       - i  w {\bf B}^{2} )} \; ,
\nonumber
\\
\gamma = {  w      -i y    \over
(x A  -  i z  B )    - (  y {\bf A} ^{2}       - i  w {\bf B}^{2} )} \; .
\label{53}
\end{eqnarray}

The formulas  allow transition to a more simple non-relativistic case
($x \equiv 0 \; ,  \; w \equiv 0 \; , \;   B = 0 $)
\begin{eqnarray}
{\bf c} = i  \; {\bf q} = i \alpha \; {\bf A} +  i \beta \; {\bf B} + i \gamma \; {\bf A} \times {\bf B} \; ,
\nonumber
\\
i \; \alpha =  -{ 1 \over {\bf A} ^{2}     } \;  {  z    \over
     y }  \; , \qquad
i \; \beta =  0 \; , \qquad
i \gamma = - {       1      \over
       {\bf A} ^{2}     } \; ;
\label{54}
\end{eqnarray}

\noindent these relations describe 1-parametric set of 3-rotations.
In relations   (\ref{48}), the non-relativistic case is reached as follow
\begin{eqnarray}
n_{0}^{2}   +   {\bf n} ^{2}    = 1   \; , \qquad
  n_{0}  =   y {\bf A} ^{2}     \; , \;\;
  {\bf n} = z   {\bf A}   + y {\bf A} \times {\bf B} \; .
\label{55}
\end{eqnarray}

let u s obtain an explicit form of the relationship
$
n_{0}^{2}  - m_{0}^{2}  = 1 + {\bf m} ^{2} -  {\bf n} ^{2} $ in  (\ref{48}).
We have
\begin{eqnarray}
n_{0}^{2}  - m_{0}^{2}   =( A \; x  -   {\bf A} ^{2}    y )^{2} - (- B \; z  +   {\bf B}^{2}\;  w)^{2}=
\nonumber
\\
= A^{2} x^{2}  -B^{2} z^{2}   - 2  A{\bf A}^{2} \; x y  +2  B{\bf B}^{2} \; z w +
 ({\bf A} ^{2})^{2} \;  y^{2} -
({\bf B} ^{2})^{2} \;  w^{2}  \; ,
\nonumber
\end{eqnarray}

\noindent and further
\begin{eqnarray}
{\bf m} ^{2} =  (x {\bf B}    - y \; B \; {\bf A}   + w {\bf A} \times {\bf B})\;
                (x {\bf B}    - y \; B \; {\bf A}   + w {\bf A} \times {\bf B})=
\nonumber
\\
x^{2} \; {\bf B}^{2} -x y \; B ({\bf B}{\bf A}) -x y \; B ({\bf B}{\bf A}) + y^{2} \; B^{2} {\bf A}^{2} +
w^{2} {\bf A}^{2} {\bf B}^{2} - w^{2} ({\bf A} {\bf B})^{2} \; ,
\nonumber
\end{eqnarray}

\noindent
that is
\begin{eqnarray}
{\bf m} ^{2} =
x^{2} \; {\bf B}^{2} - 2x y \; A B^{2}   + y^{2} \; B^{2} {\bf A}^{2} +
w^{2} {\bf A}^{2} {\bf B}^{2} - w^{2}  A^{2} B^{2}\; .
\nonumber
\end{eqnarray}

\noindent
In the same manner, we derive
\begin{eqnarray}
{\bf n}^{2}  = (z   {\bf A}   -   w  A  {\bf B}  + y {\bf A} \times {\bf B})\;
(z   {\bf A}   -   w  A  {\bf B}  + y {\bf A} \times {\bf B}) =
\nonumber
\\
= z^{2} \;  {\bf A}^{2} -2 z w \; B A^{2}  +w^{2} \; A^{2} {\bf B}^{2} +y^{2} {\bf A}^{2}{\bf B}^{2} -
y^{2} A^{2} B^{2} \; ,
\nonumber
\end{eqnarray}

\noindent
and further
\begin{eqnarray}
1 + {\bf m} ^{2} -  {\bf n} ^{2}=
1 +
x^{2} \; {\bf B}^{2} - 2x y \; A B^{2}   + y^{2} \; B^{2} {\bf A}^{2} +
w^{2} {\bf A}^{2} {\bf B}^{2} - w^{2}  A^{2} B^{2}-
\nonumber
\\
-z^{2} \;  {\bf A}^{2} + 2 z w \; B A^{2}  - w^{2} \; A^{2} {\bf B}^{2} - y^{2} {\bf A}^{2}{\bf B}^{2} +
y^{2} A^{2} B^{2} \; ,
\nonumber
\end{eqnarray}

\noindent
that is

\begin{eqnarray}
1 + {\bf m} ^{2} -  {\bf n} ^{2}=
1 +
x^{2} \; {\bf B}^{2} -
z^{2} \;  {\bf A}^{2}    - 2x y \; A B^{2} + 2 z w \; B A^{2}    +
\nonumber
\\
+  y^{2} [ ( \; B^{2}  -  {\bf B}^{2} ){\bf A}^{2}  + A^{2} B^{2} ]  -
w^{2} [ (A^{2} - {\bf A}^{2}     ) {\bf B}^{2}  +  A^{2} B^{2}  ] \; .
\nonumber
\end{eqnarray}

The quadratic equation for parameters of the  Mueller matrix
takes the form

\begin{eqnarray}
 x^{2}   (A^{2}   - {\bf B}^{2} )
       + 2x y \; A (B^{2}  -{\bf A}^{2} )
+  y^{2}  [ \; (    {\bf A}^{2} +  {\bf B}^{2} - B^{2} ){\bf A}^{2}  - A^{2} B^{2}\;  ]
=
\nonumber
\\
=    z^{2}  ( B^{2}   -  {\bf A}^{2} )        + 2 z w\;
  B (   A^{2} - {\bf B}^{2}    )   + w^{2}  [ \;
  (  {\bf A}^{2} +  {\bf B}^{2}  - A^{2}  ) {\bf B}^{2}  -  A^{2} B^{2} \; ] + 1\; .
\nonumber
\\
\label{56}
\end{eqnarray}

\subsection*{5.  On defining 4-dimensional Mueller matrix from
polarization measurements
}

As shown above, each polarization measurement
\begin{eqnarray}
S_{a} \;\; \stackrel{L}{\Longrightarrow} \;\; S_{a}'\qquad
\mbox{or}\qquad (A_{a},B_{a})  \;\; \stackrel{L}{\Longrightarrow} \;\; (A_{a}',B_{a}')
\nonumber
\end{eqnarray}

\noindent allows to obtain the quadratic  constraint
on Mueller's characteristics of  a polarization device

\begin{eqnarray}
 x^{2} \;  (A^{2}   - {\bf B}^{2} )
       + 2 xy \; A (B^{2}  -{\bf A}^{2} )
+  y^{2}\;  [ \; (    {\bf A}^{2} +  {\bf B}^{2} - B^{2} ){\bf A}^{2}  - A^{2} B^{2}\;  ]
=
\nonumber
\\
=    z^{2} \; ( B^{2}   -  {\bf A}^{2} )        + 2 zw \;
  B (   A^{2} - {\bf B}^{2}    )   + w^{2} \; [ \;
  (  {\bf A}^{2} +  {\bf B}^{2}  - A^{2}  ) {\bf B}^{2}  -  A^{2} B^{2} \; ] + 1\; ;
\nonumber
\\
\label{59}
\end{eqnarray}

\noindent the later has a 3-parametric set  of solutions which
describe all the possible  Mueler matrices of the given optical device

\begin{eqnarray}
  n_{0}  = x \; A   -   y \; {\bf A} ^{2} \; , \qquad
  {\bf n} = z  \;  {\bf A}   -   w  \; A  {\bf B}  + y \; {\bf A} \times {\bf B} \; ,
\nonumber
\\
m_{0}  = - z \; B   +   w\; {\bf B}^{2}   \; ,
\qquad
{\bf m} =  x \; {\bf B}    - y \; B \; {\bf A}   + w\;
{\bf A} \times {\bf B}  \; .
\label{60}
\end{eqnarray}

It is evident, that to fix Mueller matrix uniquely,  one should perform  several   polarization tests.
Let start with four ones -- the problem to solve is formulate as a system of 4 equations

\begin{eqnarray}
 x^{2} \;  (A^{2}_{1}   - {\bf B}^{2}_{1} )
       + 2 xy \; A_{1} (B^{2}_{1}  -{\bf A}^{2}_{1} )
+  y^{2}\;  [ \; (    {\bf A}^{2}_{1} +  {\bf B}^{2}_{1} - B^{2}_{1} ){\bf A}^{2} _{1} - A^{2}_{1} B^{2}_{1}\;  ]
=
\nonumber
\\
=    z^{2} \; ( B^{2}_{1}   -  {\bf A}^{2}_{1} )        + 2 zw \;
  B_{1} (   A^{2} _{1}- {\bf B}^{2} _{1}   )   + w^{2} \; [ \;
  (  {\bf A}^{2}_{1} +  {\bf B}^{2} _{1} - A^{2}_{1}  ) {\bf B}^{2}_{1}  -  A^{2}_{1} B^{2}_{1} \; ] + 1\; .
\nonumber
\\[3mm]
 x^{2} \;  (A^{2} _{2}  - {\bf B}^{2}_{2} )
       + 2 xy \; A _{2}(B^{2}_{2}  -{\bf A}^{2}_{2} )
+  y^{2}\;  [ \; (    {\bf A}^{2}_{2} +  {\bf B}^{2}_{2} - B^{2}_{2} ){\bf A}^{2} _{2} - A^{2} _{2}B^{2}_{2}\;  ]
=
\nonumber
\\
=    z^{2} \; ( B^{2}_{2}   -  {\bf A}^{2}_{2} )        + 2 zw \;
  B _{2}(   A^{2}_{2} - {\bf B}^{2}  _{2}  )   + w^{2} \; [ \;
  (  {\bf A}^{2}_{2} +  {\bf B}^{2} _{2} - A^{2} _{2} ) {\bf B}^{2}_{2}  -  A^{2}_{2} B^{2}_{2} \; ] + 1\; .
\nonumber
\\[3mm]
 x^{2} \;  (A^{2}_{3}   - {\bf B}^{2}_{3} )
       + 2 xy \; A_{3} (B^{2}_{3}  -{\bf A}^{2}_{3} )
+  y^{2}\;  [ \; (    {\bf A}^{2}_{3} +  {\bf B}^{2}_{3} - B^{2}_{3} ){\bf A}^{2}_{3}  - A^{2}_{3} B^{2}_{3}\;  ]
=
\nonumber
\\
=    z^{2} \; ( B^{2} _{3}  -  {\bf A}^{2}_{3} )        + 2 zw \;
  B_{3} (   A^{2}_{3} - {\bf B}^{2}_{3}    )   + w^{2} \; [ \;
  (  {\bf A}^{2}_{3} +  {\bf B}^{2}_{3}  - A^{2} _{3} ) {\bf B}^{2}_{3}  -  A^{2} B^{2}_{3} \; ] + 1\; .
\nonumber
\\[3mm]
 x^{2} \;  (A^{2} _{4}  - {\bf B}^{2}_{4} )
       + 2 xy \; A_{4} (B^{2}_{4}  -{\bf A}^{2}_{4} )
+  y^{2}\;  [ \; (    {\bf A}^{2}_{4} +  {\bf B}^{2}_{4} - B^{2}_{4} ){\bf A}^{2}_{4}  - A^{2}_{4} B^{2}_{4}\;  ]
=
\nonumber
\\
=    z^{2} \; ( B^{2}_{4}   -  {\bf A}^{2}_{4} )        + 2 zw \;
  B _{4}(   A^{2}_{4} - {\bf B}^{2} _{4}   )   + w^{2} \; [ \;
  (  {\bf A}^{2}_{4} +  {\bf B}^{2} _{4} - A^{2}_{4}  ) {\bf B}^{2}_{4}  -  A^{2}_{4} B^{2}_{4} \; ] + 1\; .
\label{61}
\end{eqnarray}

\noindent It may be presented in a symbolical  form as
\begin{eqnarray}
a_{1} x^{2} +2b_{1} xy + c_{1} y^{2} = \alpha_{1} z^{2} + 2\beta_{1} zw + \sigma_{1} w^{2} +1 \; ,
\nonumber
\\
a_{2} x^{2} +2b_{2} xy + c_{2} y^{2} = \alpha_{2} z^{2} + 2\beta_{2} zw + \sigma_{2} w^{2} +1 \; ,
\nonumber
\\
a_{1} x^{2} +2b_{3} xy + c_{3} y^{2} = \alpha_{3} z^{2} + 2\beta_{3} zw + \sigma_{3} w^{2} +1 \; ,
\nonumber
\\
a_{4} x^{2} +2b_{4} xy + c_{4} y^{2} = \alpha_{4} z^{2} + 2\beta_{4} zw + \sigma_{4} w^{2} +1 \; .
\label{62}
\end{eqnarray}

\noindent
In general, this mathematical task should have a definite  solution, though rather cumbersome one.
Indeed, we could successively  exclude the variables as follows
\begin{eqnarray}
(1) \qquad \Longrightarrow \qquad x = x(y, z,w)\; ,
\nonumber
\\
(2) \qquad  \Longrightarrow \qquad   y = y (z,w) \; , \qquad x = x( y (z,w), z,w) = \bar{x}(z,w) \; ,
\nonumber
\\
(3) \qquad  \Longrightarrow \qquad   z = z (w) \; , \qquad
(4) \qquad  \Longrightarrow \qquad   w = w (...) \; , \qquad z = z (w(...)) \; .
\nonumber
\label{63}
\end{eqnarray}

However, there exist another and more beautiful way to
solve the problem.  Indeed, let us consider 6 independent polarization measurements -- they provide us with 6
linear equations under 6 variables
\begin{eqnarray}
x^{2} \; , \;\;  y^{2} \;, \;\;  2xy\;, \qquad z^{2} \; , \;\;  w^{2} \;, \;\;  2zw\; ;
\nonumber
\end{eqnarray}
\begin{eqnarray}
a_{1} x^{2} +2b_{1} xy + c_{1} y^{2}  - \alpha_{1} z^{2} -  2\beta_{1} zw  - \sigma_{1} w^{2} =  +1 \; ,
\nonumber
\\
a_{2} x^{2} +2b_{2} xy + c_{2} y^{2} -  \alpha_{2} z^{2} -  2\beta_{2} zw -  \sigma_{2} w^{2} = +1 \; ,
\nonumber
\\
a_{1} x^{2} +2b_{3} xy + c_{3} y^{2} -  \alpha_{3} z^{2}  - 2\beta_{3} zw  - \sigma_{3} w^{2}  = +1 \; ,
\nonumber
\\
a_{4} x^{2} +2b_{4} xy + c_{4} y^{2} - \alpha_{4} z^{2} -  2\beta_{4} zw -  \sigma_{4} w^{2} = +1 \; ,
\nonumber
\\
a_{5} x^{2} +2b_{5} xy + c_{5} y^{2} - \alpha_{5} z^{2} -  2\beta_{5} zw -  \sigma_{5} w^{2} = +1 \; ,
\nonumber
\\
a_{6} x^{2} +2b_{6} xy + c_{6} y^{2} - \alpha_{6} z^{2} -  2\beta_{6} zw -  \sigma_{6} w^{2} = +1 \; .
\label{63}
\end{eqnarray}

\noindent
By physical reasons, we  cam presuppose existence of a unique solution of the
task.  This is  given by Kramer's  rule
\begin{eqnarray}
x^{2} = { \Delta_{x^{2}} \over \Delta} \; , \qquad
y^{2} = { \Delta_{y^{2}} \over \Delta} \; , \qquad
2xy = { \Delta_{2xy} \over \Delta} \; ,
\nonumber
\\
z^{2} = { \Delta_{z^{2}} \over \Delta} \; , \qquad
w^{2} = { \Delta_{w^{2}} \over \Delta} \; , \qquad
2zw = { \Delta_{2zw} \over \Delta} \; ,
\label{64}
\end{eqnarray}

\noindent
from whence it follows (evidently, arising subtleties  with $\pm$ should be examined additionally)
\begin{eqnarray}
x+y = \sqrt{
 { \Delta_{x^{2}} +  \Delta_{y^{2}} + \Delta_{2xy} \over \Delta} }\; ,
\qquad
x-y  =\sqrt{
 {  \Delta_{x^{2}} +  \Delta_{y^{2} }- \Delta_{2xy} \over \Delta } } \; ,
\nonumber
\\
z+w  = \sqrt{
 { \Delta_{z^{2}} +  \Delta_{w^{2}} + \Delta_{2zw} \over \Delta} }\; ,
\qquad
z-w  = \sqrt{
 {  \Delta_{z^{2}} +  \Delta_{w^{2} }- \Delta_{2zw} \over \Delta } }\; ,
\label{65}
\end{eqnarray}

\noindent
Recall  (see (\ref{51}) that Muller's matrices are defined by $k$-parameter
\begin{eqnarray}
k_{0} = (  x A   -i z  B )  -   (y {\bf A} ^{2}  -  i w {\bf B}^{2} ) \; ,
\nonumber
\\
{\bf k} =
  - ( y \; B   + i \;  z  ) \; {\bf A}  +
   ( x    +i    w  A ) \;  {\bf B}    +
   ( w      -i y ) {\bf A} \times {\bf B} \; ;
   \nonumber
\end{eqnarray}

\noindent
evidently, any orthogonal Lorentz matrix cannot distinguish between  $(+k_{0}, +{\bf k}) $ and
$(-k_{0}, -{\bf k}) $.

We may employ the same method in non-relativistic case as well. See  (\ref{55}); with the notation
 $z = \nu, y = N$ we have
\begin{eqnarray}
n_{0}^{2}   +   {\bf n} ^{2}    = 1   \; , \qquad
  n_{0}  =   y {\bf A} ^{2}     \; , \;\;
  {\bf n} = z   {\bf A}   + y {\bf A} \times {\bf B} \; .
\label{66}
\end{eqnarray}

\noindent Note that because
\begin{eqnarray}
{\bf A}^{2} = ( {\bf S} + {\bf S}')^{2} =  {\bf S}^{2}  + {\bf S}^{'2} + 2 {\bf S}{\bf S} =
 2 ( S^{2} + {\bf S}{\bf S} ) \; ,  \qquad {\bf A} \times {\bf B} = 2 {\bf S} \times {\bf S}' \; ,
\nonumber
\end{eqnarray}

\noindent eqs.  (\ref{66}) are equivalent to
\begin{eqnarray}
  n_{0}  =  2 y  \;  ( S^{2} + {\bf S}{\bf S} )     \; , \;\;
  {\bf n} = z   {\bf A}   + 2y \;   {\bf S} \times {\bf S}' \; .
\label{67}
\end{eqnarray}

\noindent  and thereby  coincide with (\ref{9})
\begin{eqnarray}
 n_{0} =  \beta \; ( S^{2} +
{\bf S} \; {\bf S}' ) \; ,\qquad
{\bf n} = \alpha \; ( {\bf S} +  {\bf S} '  ) + \beta \;    {\bf S} \times {\bf S}'   \; .
\label{68}
\end{eqnarray}

In this notation two independent  polarization test provide us with  a linear system
\begin{eqnarray}
y^{2} [  {\bf A}^{2}_{1}( {\bf A}^{2}_{1}  +  {\bf B}^{2}_{1} ) - ({\bf A}_{1} {\bf B}_{1})^{2}  ] +
 z^{2} {\bf A}^{2} _{1}= 1 \; ,
\nonumber
\\
y^{2} [  {\bf A}^{2}_{2}( {\bf A}^{2}_{2}  +  {\bf B}^{2}_{2} ) - ({\bf A}_{2} {\bf B}_{2})^{2}  ] +
 z^{2} {\bf A}^{2} _{2}= 1 \; ,
\label{69}
\end{eqnarray}

\noindent
its solution  is

\begin{eqnarray}
y^{2} = { ({\bf A}_{1} {\bf B}_{1})^{2} -  ({\bf A}_{2} {\bf B}_{2})^{2} \over
[  {\bf A}^{2}_{1}( {\bf A}^{2}_{1}  +  {\bf B}^{2}_{1} ) - ({\bf A}_{1} {\bf B}_{1})^{2}  ] {\bf A}_{2}^{2} -
[  {\bf A}^{2}_{2}( {\bf A}^{2}_{2}  +  {\bf B}^{2}_{2} ) - ({\bf A}_{2} {\bf B}_{2})^{2}  ] {\bf A}_{1}^{2} }\; ,
\nonumber
\\
z^{2} = { [  {\bf A}^{2}_{2}( {\bf A}^{2}_{2}  +  {\bf B}^{2}_{2} ) - ({\bf A}_{2} {\bf B}_{2})^{2}  ]-
[  {\bf A}^{2}_{1}( {\bf A}^{2}_{1}  +  {\bf B}^{2}_{1} ) - ({\bf A}_{1} {\bf B}_{1})^{2}  ] \over
[  {\bf A}^{2}_{1}( {\bf A}^{2}_{1}  +  {\bf B}^{2}_{1} ) - ({\bf A}_{1} {\bf B}_{1})^{2}  ] {\bf A}_{2}^{2} -
[  {\bf A}^{2}_{2}( {\bf A}^{2}_{2}  +  {\bf B}^{2}_{2} ) - ({\bf A}_{2} {\bf B}_{2})^{2}  ] {\bf A}_{1}^{2} }\; .
\label{70}
\end{eqnarray}

\subsection*{6. On diagonalizing the transitivity equation }

The  transitivity equation $LS = S'$ led us to a 3-surface in  4-parametric space

\begin{eqnarray}
 x^{2} \;  (A^{2}   - {\bf B}^{2} )
       + 2 xy \; A (B^{2}  -{\bf A}^{2} )
+  y^{2}\;
[ \; (    {\bf A}^{2} +  {\bf B}^{2} - B^{2} ){\bf A}^{2}  - A^{2} B^{2}\;  ] - \nonumber
\\
-    z^{2} \; ( B^{2}   -  {\bf A}^{2} )        - 2 zw \;
  B (   A^{2} - {\bf B}^{2}    )   - w^{2} \; [ \;
  (  {\bf A}^{2} +  {\bf B}^{2}  - A^{2}  ) {\bf B}^{2}  -  A^{2} B^{2} \; ] =  1\; ,
\nonumber
\\
\label{71}
\end{eqnarray}

\noindent or in symbolical form
\begin{eqnarray}
a x^{2} +2b xy + c y^{2}  - \alpha z^{2} - 2\beta zw - \sigma
w^{2} =  +1 \; .
\label{72}
\end{eqnarray}

Let us examine the possibility to transform an elementary quadratic form
to a diagonal form by mens of 3-rotation in 2-plane
\begin{eqnarray}
a x^{2} +2b xy + c y^{2} =  F X^{2} + G Y^{2} \;,
\nonumber\\
x = \cos \phi \;  X + \sin \phi \; Y  \;, \qquad
y = - \sin \phi \;X  + \cos \phi \;  Y   \;.
\label{73}
\end{eqnarray}

\noindent
Eqs.  (\ref{73}) yield
\begin{eqnarray}
a (\cos \phi \;  X + \sin \phi \; Y)^{2} +2b (\cos \phi \;  X + \sin \phi \; Y)
(- \sin \phi \;X  + \cos \phi \;  Y)   +
\nonumber
\\
+  c (- \sin \phi \;X  + \cos \phi \;  Y)^{2} =
  F X^{2} + G Y^{2} \qquad \Longrightarrow
\nonumber
\\[3mm]
a(2XY\,\sin \phi \cos \phi+X^{2}\cos^{2}\phi+Y^{2}\sin^{2}\phi)+
\nonumber
\\
+2 b[(Y^{2}-X^{2})\sin\phi\,\cos \phi+XY(\cos^{2}\phi-\sin^{2}\phi)]+
\nonumber
\\
+c(-2XY\,\sin \phi \cos \phi+X^{2}\sin^{2}\phi+Y^{2}\cos^{2}\phi)=F X^{2} + G Y^{2} \; .
\label{74}
\end{eqnarray}

\noindent So we have three equations
\begin{eqnarray}
X^{2}: \qquad a \cos^{2} \phi  -2b \sin \phi  \cos \phi  + c \sin^{2} \phi  =  F\,,
\nonumber
\\
Y^{2}: \qquad a\sin^{2}\phi+2b \sin \phi  \cos \phi+ c\cos^{2} \phi=G\,,
\nonumber
\\
2XY: \qquad a\sin \phi  \cos \phi+b(\cos^{2}\phi-\sin^{2}\phi)-c \sin \phi  \cos \phi=0\,.
\nonumber
\end{eqnarray}

\noindent With the help of the variable  $2\phi$, these are written as
\begin{eqnarray}
 a\, {\cos 2\phi+1\over 2}  -b \sin 2\phi    + c\,{1-\cos 2\phi\over 2}  =  F\,,
\nonumber
\\
 a\,{1-\cos 2\phi\over 2} +b \sin 2\phi  + c\,{\cos 2\phi+1\over 2}=G\,,
\nonumber
\\
 {a -c \over 2}\sin 2\phi  +b\cos 2\phi  =0  \; .
\label{75}
\end{eqnarray}

\noindent This results in
\begin{eqnarray}
\sin 2\phi={2b\over \sqrt{(c-a)^{2}+4b^{2}}}\,,\qquad \cos 2\phi={c-a\over \sqrt{(c-a)^{2}+4b^{2}}}\, ;
\label{76}
\end{eqnarray}

\noindent and
\begin{eqnarray}
F= {a+c\over 2}+{a-c\over 2} \,\cos 2\phi  -b \sin 2\phi     =  {a+c\over 2}-{\sqrt{(a-c)^{2}+4b^{2}}\over 2} \; ,
\nonumber
\\
G= {a+c\over 2}-{a-c\over 2} \,\cos 2\phi +b \sin 2\phi  = {a+c\over 2}+{\sqrt{(a-c)^{2}+4b^{2}}\over 2} \, .
\label{77}
\end{eqnarray}

\vspace{5mm}
In the same manner,  the second quadratic form is considered
\begin{eqnarray}
 - \alpha z^{2} - 2\beta zw - \sigma
w^{2} =  \Delta \; Z^{2} + \Gamma W^{2}
\nonumber
\\
z = \cos \rho \;  Z + \sin \rho \; W  \;, \qquad
w = - \sin \rho \;Z  + \cos \rho \;  W   \;.
\label{78}
\end{eqnarray}

\noindent For  $2\rho$ we get
\begin{eqnarray}
\sin 2\rho =
{2\beta \over \sqrt{(\sigma - \alpha )^{2}+4 \beta^{2}}}\,,
\qquad \cos 2\rho={\sigma- \alpha \over \sqrt{(\sigma-\alpha)^{2}+4 \beta^{2}}}\, ;
\label{79}
\end{eqnarray}
\begin{eqnarray}
\Delta =  {\alpha + \sigma \over 2}- {\sqrt{(\alpha - \sigma )^{2}+4 \beta^{2}}\over 2} \; ;
\nonumber
\\
\Gamma =  {\alpha + \sigma \over 2}+{\sqrt{(\alpha -\sigma )^{2}+4\beta ^{2}}\over 2} \, .
\label{80}
\end{eqnarray}

For instance, conditions at which  $F$  and $G$ are positive, and  $\Delta, \Gamma$are  negative,
are formulated in the form

\vspace{5mm}
\underline{$ (F,G,\Delta, \Gamma) \sim (+, +, - , -)$,}
\begin{eqnarray}
 a>0\; , \; c > 0\;, \qquad
 a+c > + \sqrt{(a-c)^{2}+4b^{2}}   > 0 \qquad \Longrightarrow \qquad  ac > b^{2} \; .
\nonumber
\\
 \alpha < 0\; , \; \sigma <  0\;, \qquad
 \alpha + \sigma < -  \sqrt{(\alpha -\sigma )^{2}+4 \beta^{2}}
  \qquad \Longrightarrow \qquad  \alpha \sigma  > \beta ^{2} \; .
\label{81}
\end{eqnarray}

When specifying expressions for  $a,b,c,\alpha, \beta, \sigma$
we  should distinguish between  a partly and completely polarized light.
In the case of a partly polarized and completely polarized  light  we have respectively
\begin{eqnarray}
S_{0}^{2} - {\bf S}^{2} =
S_{0}^{'2} - {\bf S}^{'2}=  0 \; , \qquad S_{0} = + \mid {\bf S} \mid \; ,
\nonumber
\\
S_{0}^{2} - {\bf S}^{2} = S_{0}^{'2} - {\bf S}^{2} > 0 , \qquad S_{0} > \mid {\bf S} \mid \; .
\nonumber
\end{eqnarray}

\noindent For  the main invariant let us use  the notation
$ S_{0}^{2} - {\bf S}^{2} =
S_{0}^{'2} - {\bf S}^{'2} =  \Sigma^{2}\; $.

Expression for $a,b, \alpha, \beta$ are given by
\begin{eqnarray}
a= (S_{0}+S_{0}')^{2} - ({\bf S}- {\bf S}')^{2} =
 2 \Sigma^{2}  + 2 ( S_{0}S_{0}' + {\bf S} {\bf S}')\; ,
\nonumber
\\
 {b \over A} =   (S_{0}-S_{0}')^{2} - ({\bf S}+ {\bf S}')^{2} =
2 \Sigma^{2}  - 2 ( S_{0}S_{0}' + {\bf S} {\bf S}') \; ,
\nonumber
\\
\alpha = (S_{0}-S_{0}')^{2} - ({\bf S}+ {\bf S}')^{2} =
 2 \Sigma^{2}  - 2 ( S_{0}S_{0}' + {\bf S} {\bf S}')\; ,
\nonumber
\\
{\beta \over B}=(S_{0}+S_{0}')^{2} - ({\bf S}- {\bf S}')^{2} =
2 \Sigma^{2}  + 2 ( S_{0}S_{0}' + {\bf S} {\bf S}') \; .
\label{82}
\end{eqnarray}

\noindent
they become simpler for a completely polarized light
\begin{eqnarray}
a_{polar} = + \;
2 ( S_{0}S_{0}' + {\bf S} {\bf S}') > 0 \; , \qquad
 {b_{polar} \over A} = - \; 2 ( S_{0}S_{0}' + {\bf S} {\bf S}')  < 0 \; ,
 \nonumber
 \\
 \alpha_{polar} = - \;
2 ( S_{0}S_{0}' + {\bf S} {\bf S}') < 0 \; , \qquad
 {\beta_{polar} \over B} = + \; 2 ( S_{0}S_{0}' + {\bf S} {\bf S}')  > 0\,.
  \label{83}
 \end{eqnarray}

Let us specify
$
  c=(    {\bf A}^{2} +  {\bf B}^{2} - B^{2} ){\bf
A}^{2}  - A^{2} B^{2}$;
accounting for
\begin{eqnarray}
 {\bf A}^{2} +  {\bf B}^{2} - B^{2} = ({\bf S}+{\bf S}')^{2} + ({\bf S} -{\bf S}')^{2} - (S_{0}-S_{0}')^{2}=
 -4\Sigma^{2} + (S_{0}+S_{0}')^{2}  \; ,
\nonumber
\\
{\bf A}^{2} = ({\bf S} + {\bf S}')^{2} \; , \qquad
A^{2} B^{2} = (S_{0}+S_{0}')^{2} (S_{0}-S_{0}')^{2}
\nonumber
\end{eqnarray}

\noindent
we get
\begin{eqnarray}
c = [  -4\Sigma^{2} + (S_{0}+S_{0}')^{2} ] ({\bf S} + {\bf S}')^{2} - (S_{0}+S_{0}')^{2} (S_{0}-S_{0}')^{2}  \;,
\nonumber
\\
c_{polar} =
 2 (S_{0}+S_{0}')^{2}   \;   ( S_{0}S_{0} + {\bf S} {\bf S}' )
 \; .
\label{84}
\end{eqnarray}

In the same mater, for     $\sigma=
  \sigma = (    {\bf B}^{2} +  {\bf A}^{2} - A^{2} ){\bf
B}^{2}  - B^{2} A^{2}  $
with relations
\begin{eqnarray}
{\bf B}^{2} +  {\bf A}^{2} - A^{2} = ({\bf S}-{\bf S}')^{2} + ({\bf S} +{\bf S}')^{2} - (S_{0}+S_{0}')^{2}=
-4\Sigma^{2} + (S_{0}-S_{0}')^{2}  \; ,
\nonumber
\\
{\bf B}^{2} = ({\bf S} - {\bf S}')^{2} \; , \qquad B^{2} A^{2} = (S_{0}-S_{0}')^{2} (S_{0}+S_{0}')^{2}
\nonumber
\end{eqnarray}

\noindent  we obtain
\begin{eqnarray}
  \sigma = [-4\Sigma^{2} + (S_{0}-S_{0}')^{2}  ] ({\bf S} - {\bf S}')^{2}- (S_{0}-S_{0}')^{2} (S_{0}+S_{0}')^{2} \; ,
  \nonumber
\\
\sigma_{polar} =
-2  (S_{0}-S_{0}')^{2}  \;  ( S_{0}S_{0} + {\bf S} {\bf S}' )
\label{85}
\end{eqnarray}

\subsection*{7.  On the Lorentz little group for a partly polarized light }

In the context op polarization optics,
some interest may have the known problem of the little Lorentz group.
What is the majority of Mueller matrices leaving invariant a given  Stokes 4-vector.
The problem is reduced to
\begin{eqnarray}
L_{b}^{\;\;a}(k, \bar{k}^{*}) \;  S _{a} = + S _{b} \; , \qquad S^{a}S_{a}  = \mbox{inv} > 0 \; ;
\label{86}
\end{eqnarray}

\noindent  with the use of a factorized form
    $L =  A\; A^{*} = A^{*}  \; A$, the previous equations are
        \begin{eqnarray}
A \; S = ( A^{*})^{-1} \; S \qquad \Longrightarrow \qquad [  A -  ( A^{*})^{-1}  \; ]\; S = 0 \; ,
\label{87}
\end{eqnarray}
\begin{eqnarray}
A =
\left |
\begin{array}{rrrr}
 k_{0}  & -k_{1}  &  -k_{2}  &  -k_{3}  \\
-k_{1} &   k_{0}  & -ik_{3}  &  ik_{2}  \\
-k_{2} &  ik_{3}  &   k_{0}  & -ik_{1}  \\
-k_{3} & -ik_{2}  &  ik_{1}  &   k_{0}
\end{array}  \right | \; , \;
 (A^{*} )^{-1}=
\left | \begin{array}{rrrr}
k^{*}_{0} & k^{*}_{1}  &  k^{*}_{2}  & k^{*}_{3}  \\
k^{*}_{1} & k^{*}_{0}  &  -ik^{*}_{3}  & ik^{*}_{2}  \\
k^{*}_{2} & ik^{*}_{3}  &  k^{*}_{0}  & -ik^{*}_{1}  \\
k^{*}_{3} & -ik^{*}_{2}  &  ik^{*}_{1}  & k^{*}_{0}
\end{array}  \right | .
\nonumber
\end{eqnarray}

\noindent
So we arrive at \begin{eqnarray}
\left |
\begin{array}{rrrr}
 (k_{0} - k_{0}^{*})  & -(k_{1} +k_{1}^{*})    &  -(k_{2} +k_{2}^{*})  &  -(k_{3} +k_{3}^{*})  \\
-(k_{1} +k_{1}^{*})   &   (k_{0} - k_{0}^{*})  & -i(k_{3} -k_{3}^{*})  &  i(k_{2} -k_{2}^{*})  \\
-(k_{2} +k_{2}^{*})   &  i(k_{3} -k_{3}^{*})   &   (k_{0} - k_{0}^{*})  & -i(k_{1} -k_{1}^{*})  \\
-(k_{3} +k_{3}^{*})   & -i (k_{2} -k_{2}^{*})  &  i (k_{1} -k_{1}^{*})  &   (k_{0} - k_{0}^{*})
\end{array}  \right |
\left | \begin{array}{c}
S_{0} \\
S_{1} \\
S_{2} \\
S_{3}
\end{array} \right | = 0
\label{88}
\end{eqnarray}

\noindent
which with notation
$
k_{0} = n_{0} + i m_{0} \;, \;   k_{j} = -i  n_{j} +  m_{j}
$
reads
\begin{eqnarray}
\left |
\begin{array}{rrrr}
i m_{0}   & -m_{1}  &  -m_{2}   &  -m_{3}  \\
-m_{1}    &   i m_{0}    & -n_{3}   &  n_{2}  \\
-m_{2}    &   n_{3}   &   i m_{0}    & - n_{1}  \\
-m_{3}     & - n_{2}  &  n_{1}  &   i m_{0}
\end{array}  \right |
\left | \begin{array}{c}
S_{0} \\
S_{1} \\
S_{2} \\
S_{3}
\end{array} \right | = 0
\label{89}
\end{eqnarray}

\noindent
Note that imposing restrictions $ m_{0}=0, m_{j}=0$,  we oftain
a more simple equation
\begin{eqnarray}
\left |
\begin{array}{rrrr}
0   &  0  &  0   &  0  \\
0    &   0    & -n_{3}   &  n_{2}  \\
0    &   n_{3}   &   0    & - n_{1}  \\
0     & - n_{2}  &  n_{1}  &   0
\end{array}  \right |
\left | \begin{array}{c}
S_{0} \\
S_{1} \\
S_{2} \\
S_{3}
\end{array} \right | = 0 \qquad \Longrightarrow \qquad {\bf n} ={  {\bf S} \over S}  \;
\label{90}
\end{eqnarray}

\noindent which describes a 1-parametric group of 3-rotations $O(\phi, {\bf n})$  about
the  axis  ${\bf S} = S {\bf n}$.
In general case, eq.  (\ref{89})  can be  presented in the vector form
\begin{eqnarray}
im_{0} S_{0} - {\bf m} {\bf S} = 0 \;, \qquad
-{\bf m} S_{0} + i m_{0} {\bf S} + {\bf n} \times {\bf  S} = 0 \; .
\label{91}
\end{eqnarray}

\noindent To have solutions in real variables, we must require   $m_{0}=0$.
Therefore, an  expression for ${\bf m}$ is
\begin{eqnarray}
 {\bf m} =
 {{\bf n} \times {\bf  S} \over S_{0}} = {\bf n} \times {\bf p} \; .
\label{92}
\end{eqnarray}

\noindent Thus, solution for the problem of little Lorentz group
is  (first it was obtained by Wigner [...])
\begin{eqnarray}
L_{b}^{\;\;a}(k, \bar{k}^{*}) \;  S _{a} = + S _{b} \; , \qquad S^{a}S_{a}  = \mbox{inv} > 0 \; ;
\nonumber
\\
k_{0} = n_{0}  + i 0  \;, \qquad   {\bf k} = -i \; {\bf  n } +  {\bf n} \times {\bf p} \; .
\; .
\label{93}
\end{eqnarray}

\noindent
Explicitly, additional condition for parameters looks
\begin{eqnarray}
k_{0}^{2} - {\bf k}^{2} = 1 \qquad \Longrightarrow \qquad
n_{0}^{2} + {\bf n}^{2} (1  - {\bf p}^{2}) + ({\bf n} {\bf p})^{2}=1 \;.
\label{94}
\end{eqnarray}

\noindent
This relationship determines a 3-parametric majority of <ueller matrices leaving invariant
the polarization vector  $S_{a}= (S_{0}, S_{0} p_{i})$ of the partly polarized light.
As known, this set of transformations  consists of a group isomorphic to $SU(2)$.

\subsection*{8. On the Lorentz little group for a completely polarized light
}

Analogous problem for a completely
polarized light
looks much the same
\begin{eqnarray}
L_{b}^{\;\;a}(k, \bar{k}^{*}) \;  S _{a} = + S _{b} \; , \qquad S^{a}S_{a}  = 0 \; ;
\nonumber
\label{98}
\end{eqnarray}

\noindent
we again have  equations
\begin{eqnarray}
im_{0} S_{0} - {\bf m} {\bf S} = 0 \; , \qquad
-{\bf m} S_{0} + i m_{0} {\bf S} + {\bf n} \times {\bf  S} = 0 \; ,
\nonumber
\end{eqnarray}

\noindent in which restriction $m_{0}=0$  must hold.
Solution  looks as follows
\begin{eqnarray}
L_{b}^{\;\;a}(k, \bar{k}^{*}) \;  S _{a} = + S _{b} \; , \qquad S^{a}S_{a}  = 0  \; ;
\nonumber
\\
k_{0} = n_{0}  + i 0  \;, \qquad   {\bf k} = -i \; {\bf  n } +  {\bf n} \times {\bf p} \;, \qquad
{\bf p}^{2}= 1\; .
\label{99}
\end{eqnarray}

\noindent
The difference arises  due to the relation
${\bf p}^{2}=1$,
\begin{eqnarray}
k_{0}^{2} - {\bf k}^{2} = 1 \qquad \Longrightarrow \qquad
n_{0}^{2} + ({\bf n} {\bf p})^{2}=1 \; .
\label{100}
\end{eqnarray}

\noindent This relationship determines a 3-parametric majority of Mueller matrices
leaving invariant a given isotropic Stokes 4-vector
$S_{a}= (S_{0}, S_{0} p_{i}), \; {\bf p}^{2}=1$.

\section*{Acknowledgements}

Author is grateful  to  participants of seminar of Laboratory of
Theoretical Physics,
 National Academy of Sciences of Belarus for discussion.

\end{document}